# The Cost-Benefit Fallacy:
# Why Cost-Benefit Analysis Is Broken and How to Fix It

By Bent Flyvbjerg and Dirk W. Bester[1]

**Abstract**

Most cost-benefit analyses assume that the estimates of costs and benefits are more or less accurate and unbiased. But what if, in reality, estimates are highly inaccurate and biased? Then the assumption that cost-benefit analysis is a rational way to improve resource allocation would be a fallacy. Based on the largest dataset of its kind, we test the assumption that cost and benefit estimates of public investments are accurate and unbiased. We find this is not the case with overwhelming statistical significance. We document the extent of cost overruns, benefit shortfalls, and forecasting bias in public investments. We further assess whether such inaccuracies seriously distort effective resource allocation, which is found to be the case. We explain our findings in behavioral terms and explore their policy implications. Finally, we conclude that cost-benefit analysis of public investments stands in need of reform and we outline four steps to such reform.

*Keywords*: Cost-benefit analysis, cost-benefit fallacy, public investment planning, forecasting, resource allocation, welfare economics, behavioral science, behavioral economics.

**Introduction**

The raison d'être of cost-benefit analysis is its contribution to effective resource allocation. This presupposes, however, that estimates of costs and benefits are more or less accurate and unbiased. What if that is not the case? That's the key question of the present paper. First, we describe the data and methodology we use to answer the question, covering cost and benefit data from 2,062 public investment projects. Second, we measure accuracy for eight investment types, including distributional information and data on statistical significance. Third, we explain our findings in behavioral terms, i.e., based on behavioral science. Finally, we conclude that cost-benefit analysis stands in need of reform, and we describe specific steps to such reform.

**Data and Methodology**

For testing the thesis that cost and benefit estimates are accurate and unbiased,[2] we collected a sample of 2,062 public investment projects with data on cost and benefit overrun. The sample includes eight investment types: Bridges, buildings, bus rapid transit (BRT), dams, power plants, rail, roads, and tunnels. Geographically, the sample incorporates investments in 104 countries on six continents, covering both developed and developing nations, with the majority of data from the United States and Europe. Historically, the data cover almost a century, from 1927 to 2013. Older investments were included to enable analyses of historical trends. For each investment in the dataset, the accuracy of cost estimates was measured by cost overrun (actual divided by estimated cost) while the accuracy of benefit estimates was measured by benefit overrun (actual divided by estimated benefits).

Data on estimated and actual costs and benefits of public investments are difficult to obtain. No statistical agency or other data service exists, from which valid and reliable data may be secured. The data therefore have to be mined item by item from the source, which is time consuming. All investments that we know of for which data on estimation accuracy were obtainable were considered for inclusion. Data were collected from a variety of sources, including annual accounts, cost and procurement accounts, revenue accounts, auditors' data, questionnaires, interviews, and other studies. Only data that could be supported by reliable documentary evidence were included, in order to avoid the well-known problems with recalled data and interviewee biases. In sum, all public investments of the eight types for which data were obtainable and were considered valid and reliable were included in the sample.

Preferably, costs and benefits would be measured over the full life-cycle of an investment. However, such data are rarely, if ever, available. Convention is therefore to measure the costs of a public investment by the proxy of construction or capital costs, and benefits by the proxy of first-year benefits. This convention is followed here. Estimated costs and benefits are the estimates made at the time of decision to build (sometimes also called the "final investment decision," or FID, based on the final business case), which is the baseline in time from which cost and benefit overrun are measured. Actual costs are measured as recorded outturn costs; actual benefits as first-year benefits, or a later value as close to this as possible, if available and if first-year benefits were not available. Estimated and actual benefits are recorded in the unit of measurement that the planners of the investment decided to use for measuring benefits. Pros and cons of measuring benefits by first-year benefits, and the issue of wider benefits, are discussed in the Annex.

Ideally, data would be available for both cost and benefit overrun for each investment in the dataset. However, data availability is far from ideal in the measurement of estimated and actual costs and benefits of public investments. For only 327 investments out of the 2,062 in the sample were data available for both cost overrun and benefit overrun. Using this ideal criterion would therefore result in scrapping large amounts of useful information for the 1,735 other investments in the sample, which would clearly be unacceptable. We therefore decided to run the statistical tests twice, first for the full sample of 2,062 investments, second for the subsample of 327 investments with data for both cost overrun and benefit overrun for each investment. This gave us the advantage of being able to test whether results are robust across different samples, which proved to be the case.

Investments were included in the sample based on data availability, as mentioned. This means that the results of the statistical analyses presented below are probably conservative, i.e., cost overruns in the investment population are most likely larger than in the sample, and benefit overruns smaller. This is because availability of data is often an indication of better-than-average investment management, and because data from badly performing investments are often not released and are therefore likely to be underrepresented in the sample. This must be kept in mind when interpreting results.

It should also be mentioned that there might be a survival or selection bias in the data. Even if cost-benefit analyses themselves were done without bias, such survival or selection bias might influence which analyses and policies get used in practice and therefore go on the record. If only policies that appear favorable on paper are implemented, there will be a selection bias towards policies that overestimate benefit and underestimate cost – even if the estimations themselves are honest and balanced. Such bias is difficult, and sometimes impossible, to deal with in statistical terms, because it is about things that did not happen, e.g., there is no ex-post outturn cost to record, and thus no cost overrun or underrun, on a policy that never happened, although an ex-ante cost estimate and cost-benefit analysis may well exist. Survival and selection bias is likely to prevail to some extent for any database that includes only policies that were undertaken and omits those that that were proposed but not undertaken, as one must for data on cost and benefit overruns. This, too, should be kept in mind when interpreting results.

Finally, it must be mentioned that the assessment of cost-benefit analysis that follows is as much about the so-called "evidence-based policy" movement as it is about cost-benefit analysis. The fundamental question asked is, "how valid and reliable is the evidence presented as justification for the public investments considered?" Three things most clearly distinguish theoretical cost-benefit analysis from practical program evaluation. First, consideration of costs, which are often not addressed in academic program evaluations. Second, consideration of multiple effects rather than one or a couple that are typically the focus of evaluation designs. And third, monetization of benefits, typically with shadow prices. The present dataset takes account of capital cost but focuses primarily on realization of predicted impacts for benefits, for reasons of data availability. This means, on the one hand, that full results are likely to be even less accurate than suggested by the reported public investment statistics, because uncertainty in shadow prices is not taken into account. On the other hand, it also means that without shadow pricing the results are as much about impact prediction as about cost-benefit analysis, something most would agree is desirable even in the absence of cost-benefit analysis. The requirements of theoretical cost-benefit analysis, including shadow pricing, are often not met in practice. This means that for large-sample research, like that reported below, the sample will inevitably include this type of more limited cost-benefit analysis/impact prediction.

Data and methodology are described further in the Annex.

**How Accurate Is Cost-Benefit Analysis?**
Table 1 and Figure 1 summarize the results from testing the thesis that cost and benefit estimates are accurate and unbiased. Taking rail as an example, average cost overrun is listed in the table as 1.40, which means that for rail investments actual costs turned out to be 40 percent higher than estimated costs, on average and in real terms, indicating substantial inaccuracy in cost estimates for rail. Average benefit overrun for rail is listed in the table as 0.66, which is evidence of a benefit shortfall of 34 percent, meaning that on average 34 percent of the estimated passengers never showed up on the actual trains, again indicating substantial inaccuracy.

[Table 1 and Figure 1 app. here]

If cost estimates were largely accurate and unbiased, cost overruns would be narrowly and more or less symmetrically distributed around 1. Eyeballing Table 1 and Figure 1 it is clear this thesis is false: average cost overrun is higher than 1 and the right tail is fat. The visual reading is verified by statistical tests that reject the thesis at an overwhelmingly high level of statistical significance (p<0.0001, Wilcoxon test).[3] While we do not expect the distribution to be normal, we still include tests of skewness (Agostino test) and kurtosis (Anscombe test) for reference. The cost overrun data has a skewness of 23.2, and the D'Agostino test confirmed this is significantly different from 0 (p<0.0001). The cost overrun data furthermore has a kurtosis of 724.1, and the Anscombe-Glynn kurtosis test found this is significantly different from 3 (p<0.0001), which is the kurtosis of a normal distribution, from which we conclude that the data have tails much fatter than normal. Cost overruns are highly inaccurate and biased for public investments, ranging from an average cost overrun for roads of 24 percent to dams at 85 percent, in real terms.[4] The fact that the data show bias is a crucial finding, because whereas errors cancel out, biases compound. Biases – and especially biases with fat tails, as here – are therefore notably worse news than errors in public investment planning.

Table 1 and Figure 1 further show that benefit estimates are also inaccurate, though less so than cost estimates. Bus rapid transit and rail investments have significant benefit shortfalls, on average respectively 58 percent and 34 percent. In contrast, for bridges, buildings, power plants, and roads benefit estimates are fairly accurate on average. If benefit estimates were generally accurate, benefit overruns would be narrowly and more or less symmetrically distributed around one. Statistical tests reject this thesis at an overwhelmingly high level of statistical significance (p<0.001, Wilcoxon test). Like cost estimates, benefit estimates are inaccurate and biased. Again the Wilcoxon test was supplemented by tests of skewness and kurtosis, and again this confirmed that the data on benefit overrun are skewed and heavy tailed. The benefit overrun data has a skewness of 1.1, and the D'Agostino test confirmed this is significantly different from 0 (p<0.0001). The benefit overrun data furthermore has a kurtosis of 6.5, and the Anscombe-Glynn kurtosis test found this is significantly different from that of a normal distribution (p<0.0001). We see that the kurtosis for cost overrun (724) is much larger than for benefit overrun (6.5), meaning that although both types of overrun are significantly more fat tailed than the normal distribution, the data on cost overrun have a more extreme tail than the data on benefit overrun, which is fat, nevertheless, as clearly documented by the left side of Figure 1. Finally, it is interesting to note a difference between the statistical and the economic significance of the benefits data. One could argue that the weighted total average benefit overrun of 0.94 is surprisingly good for economic predictions and that 0.94 is not *economically* significantly different from one, although *statistically* it is overwhelmingly significantly different (p < 0.0001). This illustrates the point that economic and statistical significance are not the same and should be considered separately. It must be remembered, however, that the samples of benefits data are small for most investment types and that further research is therefore necessary before final conclusions can be reached. It should also be mentioned that even if a difference is small, if it is mostly unidirectional and if it is repeated year in and year out, then it will compound over time and eventually might become significant, including in economic terms.

Considering cost and benefit overrun together, we see that the detected biases work in such a manner that cost overruns are not compensated by benefit overruns, but quite the opposite, on average. We also see that investment types with large average cost overruns tend to have large average benefit shortfalls. These are important findings, because not only do errors not cancel out for costs and benefits viewed separately, as documented above. On top of this, inaccuracy is accelerated by the fact that the average investment is hampered by a combination of cost overrun and benefit shortfall, undermining investment viability on two fronts, i.e., for both costs and benefits. Table 1 includes a statistical test of the thesis that cost overrun is balanced by benefit overrun, i.e., that errors of cost underestimation are compensated by similar errors of benefit underestimation (the p-values for this test are shown in the right-most column of Table 1). We see that the thesis is rejected with overwhelming statistical significance (p<0.0001, Mann-Whitney test).[5]

Finally, we tested whether error and bias in cost-benefit estimates have been reduced over time and found this is not the case. In sum, the data show that the main problem with cost-benefit analysis is not error, but bias. This is bad news, because where errors tend to cancel out each other because of their randomness, biases are systematic and therefore compound leading to results of cost-benefit analysis that are highly misleading.

It is noteworthy that for not a single of the eight investment types in Table 1 did forecasters overestimate cost and for not a single investment type did they underestimate benefits, on average. That is how strong and consistent the biases are. Theories, like Albert O. Hirschman's (2014) "Hiding Hand" and similar "just-start-digging" theories, which predict that cost underestimates will be offset by benefit underestimates of similar or larger magnitude – and that therefore resource allocation will be okay despite initial errors in estimates – are not supported by the data; they are rejected, and again at an overwhelmingly high level of statistical significance (p<0.0001) (Flyvbjerg 2016).

We see that the average investment is impaired by a combination of substantial cost underestimates compounded by significant benefit overestimates. Such systematic and significant bias in cost-benefit analysis is likely to lead to resource misallocation, including initiating investments that ultimately turn out to have negative net benefits and should never have been started, as argued by Ansar et al. (2014) and the World Bank (2010).[6] If investments are large enough and the economies where they are built are fragile, then just one major misallocation of resources in this manner, for a single investment, can negatively affect the national economy for decades, as Brazil and Pakistan have learned with their large-dam investments (Ansar et al. 2014), and Greece with the 2004 Olympic Games (Flyvbjerg and Stewart 2012).

We emphasize that the finding above – that investments are on average undermined by a combination of cost underestimates leading to cost overruns and benefit overestimates resulting in benefit shortfalls – does not mean that investments do not exist for which cost underestimates were fortuitously outweighed by similar or larger benefit underestimates. For instance, the German Karlsruhe-Bretten light-rail line, which is in the dataset, had a cost overrun in real terms of 78 percent

                                                                                                                                                             *7*

but an even larger benefit overrun of 158 percent. Similarly, the Danish Great Belt toll-road bridge – the longest suspension bridge in the world at the time of completion – had a cost overrun of 45 percent combined with a benefit overrun of 90 percent, making the investment profitable. Such investments, however, are in the minority. For the vast majority of investments (80 percent), cost overrun is not compensated by benefit overrun.

One might speculate, of course, that conceivably the one fifth of investments where benefit overrun compensated cost overrun may have generated more benefits in the aggregate than the four fifths that did not, much like just a few startups that succeed wildly may make up for the many that fail. We tested this thesis and found, at an overwhelmingly high level of statistical significance, that (a) not only is this not the case for public investments, but (b) the opposite is true since the net effect of the situation where cost overrun is not compensated by benefit overrun (80 percent of investments) is larger than the net effect of the situation where benefit overrun compensated cost overrun ($p=0.001$, two-sided Mann–Whitney test).

Table 1 shows that a typical ex-ante benefit-cost ratio produced by conventional methods is overestimated by between approximately 50 and 200 percent, depending on investment type. Again, this is a statistically significant bias showing that standard cost-benefit analysis consistently overestimates the net benefits of investments by a large margin, and therefore may not be trusted.

In addition to the standard statistical tests above, we tested results for the influence of investment type, geography, and time using Bayesian modeling. We found only few and small significant differences across investment type, geography, and time and none of them ran counter to the main conclusion that cost and benefit estimates are inaccurate, biased, and compound each other. The underestimation of costs and overestimation of benefits therefore do not seem to be driven by certain geographies or older investments. The pattern seems universal across space and time with an overwhelmingly high level of statistical support.

To conclude, the data show that cost and benefit estimates for public investments are highly inaccurate and biased. Cost underestimates and benefit overestimates are much more common than cost overestimates and benefit underestimates, at an overwhelmingly high level of statistical significance ($p<0.0001$). The findings are robust across both conventional and Bayesian statistical analysis, and across different samples, covering different sample sizes, geographies, and time periods. The average investment is impaired by a double whammy of substantial cost underestimates compounded by significant benefit overestimates. As a consequence, benefit-cost ratios are overestimated by between 50 and 200 percent, depending on investment type. The data show ex-ante cost-benefit analysis to be so misleading as to be worse than worthless, because decision makers might think they are being informed by such analysis when in fact they are being significantly misinformed about the return on planned investments. As a consequence, decision makers may give the green light to investments that should never have been started, leading to *mis*allocation of resources, where the whole purpose of cost-benefit analysis is the opposite: improved resource allocation. This points to a central problem for welfare economics, and for any type of economics or policy that relies on cost-benefit analysis: ex-ante estimates of costs and benefits are so erroneous and biased that instead of being the powerful tool for effective resource allocation and improved welfare depicted by theory, in practice cost-benefit analysis is in fact a poor and highly misleading guide for policy. In practice, biases creep in and derail the logic and good intentions of theory. To disregard this, as is common, is a fallacy. We call it the "cost-benefit fallacy" and add it to the list of other fallacies and biases in human decision making identified by behavioral science in recent years.

**The Cost-Benefit Fallacy**

We define the cost-benefit fallacy as the situation where individuals behave as if cost-benefit estimates are largely accurate and unbiased, when in fact they are highly inaccurate and biased. Two questions arise. First, how does one explain the cost-benefit fallacy in more depth? Second, how does one eliminate or mitigate the fallacy? In this section we focus on the first question, i.e., explanations. In the next section, the second, that is, cures.

The data presented above show with overwhelming statistical significance that it is not the fact that cost-benefit forecasts are in error that needs explaining, as is the common view for cost-benefit analysis, if the question of accuracy comes up at all (which it mostly does not). The fact that needs explaining is that the vast majority of cost-benefit forecasts are systematically biased, with underestimation for cost and overestimation for benefits. Our data go back 86 years and for this period the bias in cost-benefit forecasts has been constant. Cost-benefit forecasters are "predictably irrational" as regards bias, in the words of Ariely (2009). To begin to understand why, consider the following example.

Recently, the CEO of one of the biggest and most successful public infrastructure providers in the world explained to us why the cost forecast for one of their multi-billion-dollar investments – a high-speed rail line – had proven too low, resulting in significant cost overrun. There were three causes of the underestimate and overrun, according to the CEO. First, construction had taken longer than scheduled. Second, the investment, which involved extensive tunneling, had run into unexpected geological conditions, resulting in unplanned costs. Third, price inflation for both labor and materials had been higher than expected, the CEO explained. This is a typical and plausible way to account for cost overrun, which most would accept. Delays, geology, and price inflation are common stated causes of cost overrun in construction investments, as are complexity, scope changes, bad weather, archeological finds, and the like. This is the industry standard, in terms of explanation.

We analyzed the evidence for the specific investment and assured the CEO, yes, you're right. On the surface of things delays, geology, and price inflation caused your cost overrun. However, you will neither truly understand nor solve the problem as long as you see it like this, or so we argued. To really understand what happened to your investment, you would need to think in terms of root causes. And at the level of root causes, the explanation of your overrun is much simpler, with a single source: optimism. You were optimistic about the schedule, in assuming you could deliver the rail line several years faster than is usual for this type of investment, without having good reasons for this assumption. Second, you were optimistic about the geological conditions, without having investigated sufficiently. Third, you were optimistic about price variations, assuming variations would be small when in fact history shows they are large. This is optimism, pure and simple, unless you deliberately misrepresented the schedule, geology, and prices, in which case it would be strategic misrepresentation, we further argued.[7] In either case, your problem is a behavioral one, related to your own conduct and that of your organization and stakeholders. To understand and solve the problem you need behavioral science, not a better understanding of geology or market prices or better Gantt charts, or so we advised the CEO.

Behavioral scientists would agree that schedules, geology, market prices, scope changes, and complexity are relevant to understanding what goes on in public investment projects, but would not see them as root causes of inaccuracy and bias in cost-benefit forecasts. The root cause of cost overrun, according to behavioral science, is the well-documented fact that planners and managers keep underestimating the importance of schedules, geology, market prices, scope changes, and complexity in investment after investment. From the point of view of behavioral science, the mechanisms of scope changes, complex interfaces, archaeology, geology, bad weather, business cycles, etc. are not unknown to public investment planners, just as it is not unknown to planners that such mechanisms may be mitigated. However, planners often underestimate these mechanisms and overestimate the effectiveness of mitigation measures, due to well-known behavioral phenomena like overconfidence bias, the planning fallacy, and strategic misrepresentation.

In behavioral terms, scope changes etc. are manifestations of such misjudgment on the part of planners, and it is in this sense that planners' behavior is the root cause of inaccuracy and bias in cost-benefit forecasts. But because scope changes etc. are more visible than the underlying root causes, they are often mistaken for the cause of inaccuracy and bias. In behavioral terms, with scope changes as example, the causal chain starts with human bias (deliberate or not), which leads to underestimation of scope during planning which leads to unaccounted for scope changes during delivery which lead to cost overrun. Scope changes are an intermediate stage in this causal chain through which the root causes manifest themselves. Similarly with complexity, geology, and other so-called causes. With behavioral science we say to public investment planners and managers, "Your biggest risk is you!" It is not scope changes, complexity, etc. in themselves that are the main problem; it is how planners and managers misconceive and underestimate these phenomena, through overconfidence bias, the planning fallacy, strategic misrepresentation, etc. This is a profound and proven insight that behavioral science brings to cost-benefit forecasting and public investment planning, but unfortunately also an insight that is not always readily acknowledged and integrated in cost-benefit scholarship and practice. (Flyvbjerg et al. 2018).

Behavioral science entails a change of perspective: The problem with cost-benefit forecasts is not error but bias, and as long as we try to understand and solve the problem as something it is not (error), we will not solve it. Forecasts, policies, and decisions need to be de-biased, which is fundamentally different from eliminating error (Kahneman et al. 2011, Flyvbjerg 2008, 2013a). The main problem is also not cost overrun, even if overrun is what hurts and is visible and therefore gets the attention. The main problem is cost underestimation. Overrun is a consequence of underestimation, with the latter happening upstream from overrun, often years before overruns manifest. Again, if you try to understand and solve the problem as something it is not (cost overrun), you will fail. We need to solve the problem of cost underestimation to solve the problem of cost overrun.[8] Until we understand these basic insights from behavioral science, we are unlikely to get cost-benefit forecasting and resource allocation right. As long as we allow ourselves to be blinded by optimism bias, overconfidence bias, probability neglect, and other behavioral biases, we will keep reproducing the cost-benefit fallacy. The solution to overcoming the fallacy is therefore fairly straightforward: de-biasing.

**Four Steps to Cost-Benefit Reform**

Cost-benefit analysis may, no doubt, be a helpful tool in public investment policy and planning. However, cost-benefit analysis as practiced today is less than useful, because it is highly biased, as documented above. Cost-benefit theory and practice were developed long before behavioral science and have yet to adapt to its findings about the nature and causes of bias. This needs to change. We saw above that for public investments the biases are so substantial that the average cost-benefit analysis results in resource *mis*allocation, instead of contributing to the effective use of scarce assets which is the whole point of cost-benefit analysis.[9] For cost-benefit analysis to become useful in public investment policy and planning, the following needs to happen:

1. Systematic and effective de-biasing of cost-benefit forecasts.
2. Introduction of skin-in-the-game for cost-benefit forecasters.
3. Independent audits of cost-benefit forecasts.

   4. Adaption of cost-benefit forecasting to the messy, non-expert character of democratic decision making.

First, and most importantly, behavioral science predicts that any forecast – including cost-benefit estimates – is prone to bias. If forecasts are biased, they must be de-biased before they can be reliably used in policy making and investment decisions, or policies and decisions will be biased, too. Nudges will not suffice. Deliberate and precise de-biasing will be needed. Taking our clue from Kahneman and Tversky (1979), and using Kahneman's (2011: 245) so-called "outside view," we developed methods for such de-biasing for public investments. We use the distributional information about actual estimation errors in previous investments (established via ex-post studies) to precisely assess by how much the cost-benefit estimates for a planned venture must be adjusted before they may be considered effectively de-biased (Flyvbjerg et al. 2004, Flyvbjerg 2006). In his book *Thinking, Fast and Slow*, Daniel Kahneman reviews this work, and especially our advice to use empirical distributional information for de-biasing. Kahneman concludes,

> "This may be considered the single most important piece of advice regarding how to increase accuracy in forecasting [of costs and benefits] through improved methods. Using such distributional information from other ventures similar to that being forecasted is called taking an 'outside view' and is the cure to the planning fallacy [and thus to bias] ... The outside view is implemented by using a large database, which provides information on both plans and outcomes for hundreds [now thousands] of projects all over the world, and can be used to provide statistical information about the likely overruns of cost and time, and about the likely underperformance of projects of different types" (Kahneman 2011: 251-52).

The method for systematically (mathematically and statistically) taking the outside view is called "reference class forecasting" and is today used at scale in public investing around the world, including in countries where the method has been made mandatory, such as the UK and Denmark (UK Department for Transport 2006; Danish Ministry for Transport and Energy 2006, 2008). Independent evaluations of the method confirm its value and accuracy (Awojobi and Jenkins 2016, Batselier 2016, Batselier and Vanhoucke 2017, Bordley 2014, Chang et al. 2016, Kim et al. 2011, Liu and Napier 2010, and Liu et al. 2010).[10] Only with such de-biasing will cost-benefit forecasts be accurate and help allocate resources effectively.

Second, better incentives for accuracy in cost-benefit forecasts should be introduced. Better methods alone will not solve the problem. Akerlof and Shiller (2009: 146) suggest "firing the forecaster" when forecasts are very wrong and the consequences severe. Flyvbjerg (2013: 771-772) goes one step further in proposing "suing the forecaster" in cases of gross neglect and gross deliberate manipulation of forecasts and investors. The first court cases for public investment forecasting were decided in 2015, for Sydney's Lane Cove toll tunnel, which went bankrupt when more than half the forecasted cars never appeared in the tunnel, and for Brisbane's Clem 7 tunnel and Airport Link, which had comparable problems (Rubin 2017, Saulwick 2014, Worthington 2012). More litigation quickly followed in the United States, treating traffic forecasters as criminals (Dezember and Glazer 2013, Evans 2010, Hals 2013, Miller 2013, Singh 2017, Wright 2014). Recently, the judge responsible for the Muskrat Falls dam inquiry in Canada decided to report executives to the police for possible criminal charges related to cost underestimation for the dam (LeBlanc 2020, Roberts 2020). Similarly, executives who misled investors about the completion date and costs for two nuclear reactors at the V.C. Summer plant in South Carolina were prosecuted and pleaded guilty to felony fraud charges in federal and state court (Collins 2020). Criminalizing planners and cost-benefit analysts like this has sent shock-waves through the global forecasting and cost-benefit community, contributing to much-needed discipline and accountability. Here we suggest, as a more general and more moderate heuristic, that often it would make sense to consider giving cost-benefit forecasters skin in the game. Lawmakers and policymakers should develop institutional setups that reward forecasters who get their estimates right and punish those who do not. We should not be surprised that cost-benefit estimates are wrong if forecasters have no incentive to get them right. True, forecasters are supposed to be neutral and unbiased. But to confuse such normativity with reality is naive, for policymakers and scholars alike.

Third, independent audits are needed to ensure that points one and two above − better methods and better incentives for more accurate forecasts − work according to intention, and to adjust them when they do not. Auditing must be effectively separated from political motivations, to avoid the type of political distortion of cost-benefit forecasts described in detail by the World Bank (2010), Harrington (2006), Flyvbjerg et al. (2002, 2004), Harrington et al. (2000), and Wachs (1986, 1989, 1990, 2013). If this does not happen, audits will not and cannot serve their purpose. Audits should allow for transparency and external scrutiny, including scrutiny by independent experts, the public, and media.

Finally, and perhaps most importantly, for cost-benefit analysis to be accepted and have impact it must be understood and practiced *not* in the top-down, technocratic fashion that is common historically but, instead, in ways that fit with the messy, non-expert character of present-day democratic decision making. Here cost-benefit analysis is just one of many inputs that are amalgamated in the overall decision-making process. Top-down, technocratic approaches are anachronistic and will not work in this context. Kahneman (2011: 141-42) contrasts cost-benefit *technocrats* with cost-benefit *democrats*, exemplifying technocrats with Harvard law professor Cass Sunstein and democrats with University of Oregon psychology professor Paul Slovic. Kahneman writes about Sunstein, who is a self-declared cost-benefit technocrat and whose work Kahneman otherwise holds in high regard:

> "He [Sunstein] starts from the position that risk regulation and government intervention to reduce risks should be guided by rational weighting of costs and benefits ... he has faith in the objectivity that may be achieved by science, expertise, and careful deliberation ... Cass Sunstein would seek mechanisms [like the cost-benefit principle] that insulate decision makers from public pressures, letting the allocation of resources be determined by impartial experts who have a broad view of all risks and of the resources available to reduce them."

Kahneman here captures the key beliefs and a priori assumptions of cost-benefit technocrats, including (a) the view that cost-benefit analysis is a force for good in deciding policies; (b) faith in the objectivity of science as a basis for cost-benefit calculations; and (c) belief in impartial experts for making better decisions. Kahneman (2011: 140-45) juxtaposes this position with that of Slovic and other cost-benefit democrats, who Kahneman says "trusts the experts much less and the public somewhat more than Sunstein does." Kahneman (p. 144) approvingly cites Slovic (2000) for arguing that expert-based decision making, like that supported by cost-benefit technocrats, produces policies that the public will reject, which is an untenable situation in a democracy, according to Kahneman. "Slovic rightly stresses the resistance of the public to the idea of decisions being made by unelected and unaccountable experts," concludes Kahneman (2011: 144) with Slovic.

Kahneman (2011: 144) considers both Sunstein and Slovic "eminently sensible" and he explicitly states, "I agree with both." We agree with Kahneman. Nobody in their right mind would argue that expert input, science, and cost-benefit analysis are undesirable in deciding which policies and investments to pursue in a society. But, similarly, no one in their right mind would depend on only experts in making such decisions, especially when we know that experts are subject to the cost-benefit fallacy and produce, over and over, the kind of biased cost-benefit analyses documented above. Sunstein and other cost-benefit technocrats overemphasize the importance of experts and place too much trust in them. By supplementing Sunstein with Slovic – technocracy with democracy – Kahneman (2011: 144-45) develops a more balanced position, from which he argues that democracy "is inevitably messy" and emphasizes that the public's concerns, "even if they are unreasonable, should not be ignored by policy makers." Experts, science, and cost-benefit analysis are not enough to arrive at high-quality policy decisions, argues Kahneman. The messy process of democracy is called for to merge all concerns, including unreasonable ones, because only in this manner will policy be able to muster the support necessary to be successful.

The four measures for better cost-benefit analysis outlined above are not ivory tower theory. They have recently been implemented in innovative, full-scale experiments around the world, with encouraging results, as documented by Kahneman (2011: 249-53), Flyvbjerg et al. (2016, 2018), HM Treasury (2003), the UK Department for Transport (2006), the Danish Ministry for Transport and Energy (2006, 2008), and the Swiss Association of Road and Transportation Experts (2006). More such work is in the pipeline, bringing the findings of behavioral science to cost-benefit forecasting, resulting in more realistic and more useful forecasts.

**Summary and Conclusions**

In this paper, we set out to test the accuracy of cost-benefit analysis of public investments. We found a fallacy at the heart of conventional cost-benefit analysis: forecasters, policymakers, and scholars tend to assume that cost-benefit forecasts are more or less accurate, when in fact they are highly inaccurate and biased, at an overwhelmingly high level of statistical significance. In public capital investing, the fallacy results in average overestimates in ex-ante benefit-cost ratios of 50 to 200 percent, depending on investment type. Individual estimates of benefit-cost ratios are routinely off by much more than these averages.

Cost-benefit analysis can undoubtedly be a useful tool in public investment policy and practice, but only after the cost-benefit fallacy has been acknowledged and corrected for. For this to happen, the following must become part and parcel of cost-benefit practice: (a) new methods for effective de-biasing of ex-ante cost-benefit estimates, based on behavioral science; (b) incentives for using these methods, including skin-in-the-game; (c) independent audits to check that the methods and incentives work according to intention; and, finally, (d) integration of results in messy, real-life democratic decision-making processes, which consider other concerns than the results of cost-benefit analyses. The good news is that the theoretical and methodological rationale for this type of behaviorally based cost-benefit analysis has already been developed and is finding use in policy and practice. The bad news is that the cost-benefit fallacy seems to be as ingrained in human behavior as the many other biases identified by behavioral science and will therefore be as difficult to root out. That should not keep us from trying, needless to say, which is what we have done with the research reported in the present paper.

**Acknowledgments**

The authors wish to thank the Editor-in-Chief and four anonymous reviewers for their excellent insights and penetrating comments, which greatly helped improve the paper.

**Annex: Sample and Statistical Tests**

The sample for testing accuracy and bias in cost-benefit analysis includes 2,062 public investment projects across eight investment types: Bridges, buildings, bus rapid transit (BRT), dams, power plants, rail, roads, and tunnels. Geographically, the investments are located in 104 countries on six continents, including both developed and developing nations. Historically, the data cover the period from 1927 to 2013. Older investments were included to enable analyses of historical trends. For each investment in the dataset, the accuracy of cost estimates was measured by cost overrun (actual divided by estimated cost, in real terms) while the accuracy of benefit estimates was measured by benefit overrun (actual divided by estimated benefits).

Data on estimated and actual costs and benefits of public investments are difficult to obtain. No statistical agency or other data service exists, from which valid and reliable data may be secured. Over more than 20 years, the authors and their associates have therefore developed their own dataset, the largest of its kind. Data collection began with public investments in transportation, following the methodology pioneered by Pickrell (1990) in his classic study of forecast versus actual cost and utilization in US urban rail transit investments. Pickrell systematized ideas first employed by Hamer (1976). Pickrell rigorously measured estimated and actual costs and benefits in ten transit investments and insisted that the estimates to be compared to actual outcomes were those on the basis of which planners and officials had decided to proceed with proposed investments. Pickrell argued that revisionist forecasts issued by planners and officials later in the planning and tendering process or after construction had commenced – in short, after an irrevocable commitment to proceed with the investment had been made – were irrelevant, and comparing them to subsequent actual outcomes, as was (and is) common, was misleading, because later forecasts did not influence the decision to proceed with the investment. Pickrell's methodology has become an international standard for collecting and comparing estimated and actual costs and benefits in transportation capital investments and beyond.

Our first dataset following this standard was published in Flyvbjerg et al. (2002, 2005), covering 258 transportation capital investments. This was the first time an academic dataset was large enough to allow statistically valid conclusions regarding the accuracy of cost and benefit estimates in transportation capital investments. Since then, we have expanded the dataset to cover more transportation investments and to also include other types of public investments, e.g., dams, power plants, and buildings. For two decades, we have continuously monitored and grown the dataset from two sources. First, we collected our own data directly from public investments for which data were available when the investments were decided (estimated costs and benefits) *and* later completed (actual costs and benefits). Second, we included data for which other researchers in other studies had done the data collection, to allow meta-analysis. For both sources, we used formal and informal search. We did (and do) formal bibliographic searches on a continuous basis using Google Scholar, Web of Science, and individual journals, with keywords like "cost-benefit analysis," "project costs," "project benefits," "(in)accuracy of cost-benefit analysis," "cost overrun," "cost overestimate," "cost underestimate," "benefit shortfall," "benefit overrun," "benefit overestimate," and "benefit underestimate." In addition, our professional network informally kept (and keeps) us informed about new investments and new studies with relevant data. Finally, after publication of our 2002 and 2005 studies, these and later research became so widely cited that we were able to reliably discover new studies with new data simply by tracking who cited our publications, which therefore became a highly reliable part of our systematic bibliographic searches. For increased accuracy, we triangulated the formal and informal searches against each other. We deem that all public investments and all studies for which data on estimated and actual costs and benefits are publicly available have been considered for inclusion in our dataset, for the investment types covered by the study. On that basis, we consider it unlikely that important data exist that are not included, or were not considered for inclusion.

Whether data were collected by us or by other researchers they typically came from annual accounts, cost and procurement records, revenue accounts, auditors' data, questionnaires, or interviews. Only data that could be supported by reliable documentary evidence were included, in order to avoid the well-known problems with recalled data and interviewee biases. Even so, substantial amounts of data had to be rejected due to insufficient data quality, approximately 25 percent of cost data and 50 percent of benefit data. Common reasons for data rejection were lack of clarity regarding (a) baselines (not knowing the year for which data apply and from which accuracy was calculated), (b) price levels (not knowing in which year's prices monetary values were calculated), (c) exchange rates (not knowing the basket of exchange rates used to convert local currencies into US dollars), and (d) discount rates and price indices (not knowing how monetary values were conflated over time). Data were also rejected for which outliers had been omitted for no good reason, or for which such omission was clearly statistically unsound (see further Flyvbjerg et al. 2018). For studies done by others, each data point was checked for validity and reliability as well as for duplication to avoid double counting. If data were already included in our dataset, or if they were not valid and reliable, they were left out.

Finally, it should be mentioned that recently it was discovered that some ex-post studies of cost forecasting have been contaminated by significant statistical errors, for instance in pooling data with incompatible baselines and excluding outliers that should have been included. Such studies cannot be trusted. The same holds for meta-studies that included these studies without being aware of their shortcomings. This unhappy situation – which signifies the arrival of junk science in public investment scholarship – is accounted for in Flyvbjerg et al. (2018, 2019) together with recommendations on how to avoid the problem going forward. The importance of rooting out studies with faulty data and statistics from the body of work in public investment scholarship cannot be overemphasized, if the field wants to enjoy continued credence in the academy, policy making, and with the general public. Data from studies that proved contaminated in this manner are not included in the present research, needless to say.

In sum, following the methodology described above all investments for which data were considered valid and reliable were included in the present study, a total of 2,062 investments. Data collection and the dataset are described in more detail in Flyvbjerg et al. (2002, 2005) and Flyvbjerg (2016).

Preferably, costs and benefits would be measured over the full life-cycle of an investment. However, such data are rarely, if ever, available. Convention is therefore to measure the cost of investments by the proxy of construction or capital costs, and benefits by the proxy of first-year benefits. This convention is followed here. Estimated costs and benefits are the estimates made at the time of decision to build (sometimes also called the "final investment decision," or FID, based on the final business case), which is the baseline in time from which cost and benefit overrun are measured. Actual costs are measured as recorded outturn costs; actual benefits as first-year benefits, or a later value as close to this as possible, if available and if first-year benefits were not available. Estimated and actual benefits are recorded in the unit of measurement that the planners of the investment decided to use for measuring benefits.

First-year benefits may seem a narrow proxy to use for benefits, and it has been criticized as such. In fact, however, first-year benefits have proven a reliable proxy for overall benefits, which is fortunate because the existence of benefits data for later years is so rare that benefits measurement would be rendered impossible for large samples if one had to rely on later-year data. For public investments for which data are available on estimated and actual benefits covering more than one year after operations began, it turns out that investments with lower-than-estimated benefits during the first year of operations also tend to have lower-than-estimated benefits in later years (Flyvbjerg 2013: 766–767). Using the first year as the basis for measuring benefits therefore appears not often to result in the error of identifying investments as underperforming in terms of benefits that would not be identified as such if a different time period were used as the basis for comparison. Actual benefits do not seem to quickly catch up with estimated benefits for investments with overestimated first-year benefits, and mostly they never do. Ramp up of benefits is commonly assumed, but often does not happen, or occurs only partly. To take a typical example, for the Channel tunnel between the UK and France, more than five years after opening to the public, train passengers numbered only 45 percent of that forecasted for the opening year; rail freight traffic was 40 percent of the forecast; and actual numbers had not caught up with the forecasts after 20 years, and never have. In conclusion, we would prefer to measure benefits for all years, but data for this are generally unavailable. Given this is the case, the good news is that first-year benefits appear to be a fair proxy, with data more readily available than for later years.

Using first-year benefits has also been criticized for not considering wider development effects, for instance increased real estate values following from improved transport services. Such effects undoubtedly exist, but if transport investments have wider benefits, these must be expected to be roughly proportional to traffic, so that if traffic has been overestimated, so must have the development benefits, which means that the case for proceeding with the investment was exaggerated. The difference between estimated and actual traffic, including first-year patronage, would therefore be a good proxy for assessing the extent of the problem. The same applies to other types of investments. For the full argument and for further documentation regarding wider benefits, see Vickerman (2017) and Flyvbjerg (2005, 2013).

Ideally, data would be available for both cost and benefit overrun for each investment in the dataset. However, data availability is far from ideal in the measurement of estimated and actual costs and benefits of public investments. For only 327 investments out of the 2,062 in the sample were data available for both cost overrun and benefit overrun. Using this ideal criterion would therefore result in scrapping large amounts of useful information for the 1,735 other investments in the sample, which would clearly be unacceptable. We therefore decided to run the statistical tests twice, first for the full sample of 2,062 investments, second for the subsample of 327 investments with data available for both cost overrun and benefit overrun for each investment. This gave us the advantage of being able to test whether results are robust across different samples, which proved to be the case.

Investments were included in the sample based on data availability, as mentioned. This means that the results of statistical analyses presented in the main text are probably conservative, i.e., cost overruns in the investment population are most likely larger than in the sample, and benefit overruns smaller. This is because availability of data is often an indication of better-than-average investment management, and because data from badly performing investments are often not released and are therefore likely to be underrepresented in the sample. This must be kept in mind when interpreting results. See further Flyvbjerg et al. (2002, 2005) and Flyvbjerg (2005, 2016).

Standard statistical tests are described in the main text. In addition to these, we tested results for the influence of investment type, geography, and time using Bayesian modeling.[11] We found only few significant differences across investment type, geography, and time and none of them ran counter to the main conclusion that cost and benefit estimates are inaccurate, biased, and compound each other. This is unsurprising, given the overwhelmingly high level of statistical support for the main conclusion. In sum, results from the statistical tests proved robust across both conventional and Bayesian testing, and across different samples.

*Table 1: Cost-benefit estimates are inaccurate, biased, and compound each other. Accuracy is measured by cost and benefit overruns in 2,062 public investments. Cost and benefit overruns are measured as actual divided by estimated costs and benefits (A/E), respectively, in real terms.*

| Investment type | Cost overrun (A/E) | | | Benefit overrun (A/E) | | | p** |
|---|---|---|---|---|---|---|---|
| | N | Average | p* | N | Average | p* | |
| Dams | 243 | 1.96 | < 0.0001 | 84 | 0.89 | < 0.0001 | <0.0001 |
| BRT† | 6 | 1.41 | 0.031 | 4 | 0.42 | 0.12 | 0.007 |
| Rail | 264 | 1.40 | < 0.0001 | 74 | 0.66 | < 0.0001 | <0.0001 |
| Tunnels | 48 | 1.36 | < 0.0001 | 23 | 0.81 | 0.03 | 0.015 |
| Power plants | 100 | 1.36 | 0.0076 | 23 | 0.94 | 0.11 | 0.0003 |
| Buildings | 24 | 1.36 | 0.00087 | 20 | 0.99 | 0.77 | 0.01 |
| Bridges | 49 | 1.32 | 0.00012 | 26 | 0.96 | 0.099 | <0.0001 |
| Roads | 869 | 1.24 | < 0.0001 | 532 | 0.96 | < 0.0001 | <0.0001 |
| **Total** | **1603** | **1.39/1.43††** | **< 0.0001** | **786** | **0.94/0.83††** | **< 0.0001** | **<0.0001** |

\*) The p-value of Wilcoxon test with null hypothesis that the distribution is symmetrically centered around one.

\*\*) The p-value of the test with null hypothesis that cost overrun is balanced by benefit overrun (Mann-Whitney test). See main text for explanation.

†) Bus rapid transit.

††) Weighted and unweighted average, respectively.

*Figure 1: Two histograms showing the relative frequency of cost and benefit overruns, respectively, pooling all investments (N = 2,062). Values larger than or equal to 3.0 have been lumped together. If estimates of costs and benefits were largely accurate and unbiased, both histograms would be more or less symmetrically distributed around 1 with thin tails (technically, we would expect the log of the distribution to be exactly symmetric around one). In fact, each histogram is highly asymmetrical, not distributed around 1, and fat tailed (to the left for benefits, to the right for costs). On average, cost overruns are compounded by benefit shortfalls instead of being mitigated by benefit overruns. The findings are overwhelmingly statistically significant (p < 0.0001).*

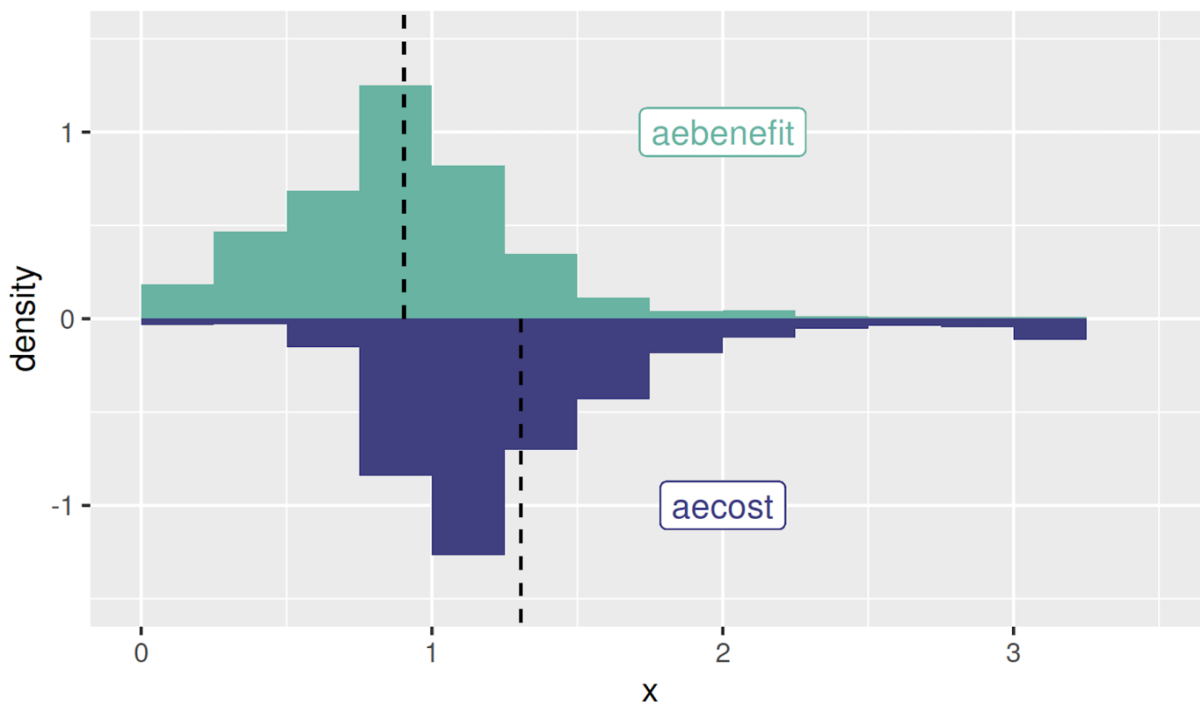

aebenefit: actual benefit divided by estimated benefit.
aecost: actual cost divided by estimated cost.
≥ 3.0: values have been pooled.
-----: average.

**Endnotes**

[1] Bent Flyvbjerg is the first BT Professor and Inaugural Chair of Major Programme Management at the University of Oxford and the Villum Kann Rasmussen Professor and Chair of Major Program Management at the IT University of Copenhagen. Dirk W. Bester holds his DPhil in statistics from the Department of Statistics at the University of Oxford and works as a professional statistician in research and industry. Flyvbjerg developed the ideas for the paper, did the research, and authored the text. Bester developed the statistical proofs and did all statistical analyses. Flyvbjerg is corresponding author (flyvbjerg@mac.com).

[2] Here and throughout the paper, we use the term "bias" in the sense it is used in behavioral science, including behavioral economics. This should not be confused with the way the term is used in statistics, in the frequentist-estimator sense, especially for the statistical tests reported in the main text and the annex.

[3] Significance is here defined in the conventional manner, with $p \leq 0.05$ being significant, $p \leq 0.01$ very significant, and $p \leq 0.001$ overwhelmingly significant.

[4] These findings correspond to other findings based on smaller samples, which we take to attest to the robustness of results (see, e.g., Albalate and Bel 2014; Altshuler and Luberoff, 2003; Ansar et al. 2014; Bain 2009; Bain and Wilkins 2002; Cantarelli et al. 2012; Dantata et al. 2006; Federal Transit Administration 2003, 2008, 2013; Flyvbjerg et al. 2002, 2004, 2005; Fouracre et al. 1990; Gao and Touran 2020; Huo et al. 2018, Leavitt et al. 1993; Lee 2008; National Audit Office 1992; Nijkamp and Ubbels 1999; Pickrell 1990, 1992; Riksrevisionen 2011; Riksrevisionsverket 1994; Walmsley and Pickett 1992; World Bank 1994).

[5] We ran the same tests with similar results for a subsample of 327 investments for which data were available for both cost overrun and benefit overrun for each investment. See further Annex 1.

[6] Negative impact on GDP is not limited to public investments. Misallocation of resources similarly happens in the private sector with similar consequences. For private-sector examples, see Flyvbjerg and Budzier (2011).

[7] Strategic misrepresentation is deliberate, while optimism is non-deliberate. See more in Flyvbjerg (2007).

[8] Cost overrun may also originate with subpar delivery, needless to say. However, if the original cost estimate for an investment is truly optimistic, as is common, then no matter how good the delivery team is they will not be able to deliver to the estimate (Flyvbjerg 2016).

[9] For an in-depth case study of how cost-benefit analysis may lead to resource misallocation, even in an organization with deep experience with and extensive use of the method, see the World Bank (2010).

[10] A typical assessment of reference class forecasting (RCF) concludes: "The conducted evaluation is entirely based on real-life project data and shows that RCF indeed performs best, for both cost and time forecasting, and therefore supports the practical relevance of the technique" (Batselier and Vanhoucke 2016: 36).

[11] Parameters for the Bayesian models were estimated using MCMC. The language JAGS was used for this, through the rjags interface to R (Plummer, 2003, 2012; R Core Team, 2012). Statistical significance for the Bayesian tests was measured by the Bayes Factor (BF) instead of by p-values, where $12 < BF \leq 150$ indicates a statistically significant result and $BF > 150$ indicates a highly significant result.